\documentclass[11pt,english,a4paper, 11pt]{article}
\usepackage[T1]{fontenc}
\usepackage[latin9]{inputenc}
\usepackage{mathtools}
\usepackage{amsmath}
\usepackage{microtype}

\makeatletter
\usepackage{jheppub} 
\toccontinuoustrue
\usepackage{microtype}

\makeatother

\usepackage{babel}
\begin{document}
\title{Fifteen-vertex models with non-symmetric \boldmath $R$ matrices}

\affiliation[a]{Departamento de Matemática, Universidade Federal de São Carlos (UFSCar), \\ Rod. Washington Luís, Km 235, C.P. 676, CEP. 13565-905, São Carlos, SP, Brasil}

\author[a]{R. S. Vieira}

\emailAdd{rsvieira@dm.ufscar.br}

\abstract{In this work, we employ the algebraic-differential method recently developed by the author to solve the Yang-Baxter equation for arbitrary fifteen-vertex models satisfying the ice-rule. We show that there are four different families of such regular $R$ matrices containing several free-parameters. The corresponding reflection $K$ matrices, solutions of the boundary Yang-Baxter equation, were also found and classified. We found that there are three different families of regular $K$ matrices, regardless of what $R$ matrix we choose.}

\keywords{Yang-Baxter equation, Bethe Ansatz, Differential and Algebraic Geometry, Lattice Integrable Models}

\maketitle

\section{The \boldmath $R$ matrices}

In this work, we classify all the regular solutions of the Yang-Baxter
equation \cite{yang1967some,Baxter1972partition}, 
\begin{equation}
R_{12}\left(u\right)R_{23}\left(u+v\right)R_{13}\left(v\right)=R_{13}\left(v\right)R_{23}\left(u+v\right)R_{12}\left(u\right),\label{YBE}
\end{equation}
for fifteen-vertex models whose $R$ matrix satisfies the ice-rule
and has the standard form (hereafter we shall omit the dependence
of the $R$ matrix elements on the spectral parameter $u$): 
\begin{equation}
R=\begin{pmatrix}r_{11} & 0 & 0 & 0 & 0 & 0 & 0 & 0 & 0\\
0 & r_{22} & 0 & r_{24} & 0 & 0 & 0 & 0 & 0\\
0 & 0 & r_{33} & 0 & 0 & 0 & r_{37} & 0 & 0\\
0 & r_{42} & 0 & r_{44} & 0 & 0 & 0 & 0 & 0\\
0 & 0 & 0 & 0 & r_{55} & 0 & 0 & 0 & 0\\
0 & 0 & 0 & 0 & 0 & r_{66} & 0 & r_{68} & 0\\
0 & 0 & r_{73} & 0 & 0 & 0 & r_{77} & 0 & 0\\
0 & 0 & 0 & 0 & 0 & r_{86} & 0 & r_{88} & 0\\
0 & 0 & 0 & 0 & 0 & 0 & 0 & 0 & r_{99}
\end{pmatrix}.\label{R}
\end{equation}

The fifteen-vertex $R$ matrices reported here were found through
an algebraic-differential method first introduced in \cite{vieira2018solving}.
The details of the computations are described in Appendix \ref{AppR}.
This represents the first step towards the classification of the most
general, \emph{non-symmetric}, $R$ matrices associated with spin-1
(three-states) vertex models. Spin-1 vertex models include important
models of statistical mechanics as, for example, the fifteen-vertex
models due to Cherednik, Babelon, Perk and Schultz \cite{cherednik1980method,babelon1981solutions,perk1981new},\emph{
t-J} models \cite{schlottmann1987integrable}, the nineteen-vertex
models of Zamolodchikov-Fateev \cite{zamolodchikov1980model} and
Izergin-Korepin \cite{izergin1981inverse}, among other graded (supersymmetric)
models. In \cite{idzumi1994solvable}, Idzumi et al. proposed a classification
of the spin-1 vertex models satisfying the ice-rule (whose most general
$R$ matrix corresponds to that one of the nineteen-vertex model).
However, the authors of \cite{idzumi1994solvable} assumed many symmetries
for the $R$ matrices from the very start, so that only the \emph{symmetric}
solutions were actually classified. 

We found that there exist four families of regular $R$ matrices for
such fifteen-vertex models. All the four solutions share the following
common part: 
\begin{align}
r_{24} & =\mathrm{e}^{\alpha_{24}u}, & r_{42} & =\mathrm{e}^{\alpha_{42}u},\nonumber \\
r_{37} & =\mathrm{e}^{\alpha_{37}u}, & r_{73} & =\mathrm{e}^{\left(\alpha_{24}-\alpha_{37}+\alpha_{42}\right)u},\nonumber \\
r_{68} & =\mathrm{e}^{\alpha_{68}u}, & r_{86} & =\mathrm{e}^{\left(\alpha_{24}+\alpha_{42}-\alpha_{68}\right)u},\nonumber \\
r_{22} & =\frac{\alpha_{22}}{\omega}\mathrm{e}^{\frac{1}{2}\left(\alpha_{24}+\alpha_{42}\right)u}\sinh\left(\omega u\right), & r_{44} & =\frac{\varOmega}{\alpha_{22}}\mathrm{e}^{\frac{1}{2}\left(\alpha_{24}+\alpha_{42}\right)u}\sinh\left(\omega u\right),\nonumber \\
r_{33} & =\frac{\alpha_{33}}{\omega}\mathrm{e}^{\frac{1}{2}\left(\alpha_{24}+\alpha_{42}\right)u}\sinh\left(\omega u\right), & r_{77} & =\frac{\varOmega}{\alpha_{33}}\mathrm{e}^{\frac{1}{2}\left(\alpha_{24}+\alpha_{42}\right)u}\sinh\left(\omega u\right),\nonumber \\
r_{66} & =\frac{\alpha_{66}}{\omega}\mathrm{e}^{\frac{1}{2}\left(\alpha_{24}+\alpha_{42}\right)u}\sinh\left(\omega u\right), & r_{88} & =\frac{\varOmega}{\alpha_{66}}\mathrm{e}^{\frac{1}{2}\left(\alpha_{24}+\alpha_{42}\right)u}\sinh\left(\omega u\right),
\end{align}
 where we introduced the quantities:
\begin{align}
\omega & =\frac{1}{2}\left(\alpha_{24}-2\alpha_{37}-\alpha_{42}+2\alpha_{68}\right),\nonumber \\
\varOmega & =\frac{1}{\omega}\left(\alpha_{11}-\alpha_{24}+\alpha_{37}-\alpha_{68}\right)\left(\alpha_{11}-\alpha_{37}-\alpha_{42}+\alpha_{68}\right).\label{Omegas}
\end{align}

The four solutions differ from each other only on the expressions
for the diagonal elements $r_{11}$, $r_{55}$ and $r_{99}$. In fact,
for the first solution, we have,
\begin{equation}
r_{11}=r_{55}=r_{99}=\mathrm{e}^{\frac{1}{2}\left(\alpha_{24}+\alpha_{42}\right)u}\frac{\sinh\left[\omega\left(\eta+u\right)\right]}{\sinh\left(\omega\eta\right)},
\end{equation}
while, for the second solution, we have,
\begin{align}
r_{11} & =r_{55}=\mathrm{e}^{\frac{1}{2}\left(\alpha_{24}+\alpha_{42}\right)u}\frac{\sinh\left[\omega\left(\eta+u\right)\right]}{\sinh\left(\omega\eta\right)}, & r_{99} & =\mathrm{e}^{\frac{1}{2}\left(\alpha_{24}+\alpha_{42}\right)u}\frac{\sinh\left[\omega\left(\eta-u\right)\right]}{\sinh\left(\omega\eta\right)}.
\end{align}
For the third solution, we get,
\begin{align}
r_{11} & =r_{99}=\mathrm{e}^{\frac{1}{2}\left(\alpha_{24}+\alpha_{42}\right)u}\frac{\sinh\left[\omega\left(\eta+u\right)\right]}{\sinh\left(\omega\eta\right)}, & r_{55} & =\mathrm{e}^{\frac{1}{2}\left(\alpha_{24}+\alpha_{42}\right)u}\frac{\sinh\left[\omega\left(\eta-u\right)\right]}{\sinh\left(\omega\eta\right)},
\end{align}
and, finally, for the fourth solution, we have, 
\begin{align}
r_{11} & =\mathrm{e}^{\frac{1}{2}\left(\alpha_{24}+\alpha_{42}\right)u}\frac{\sinh\left[\omega\left(\eta+u\right)\right]}{\sinh\left(\omega\eta\right)}, & r_{55} & =r_{99}=\mathrm{e}^{\frac{1}{2}\left(\alpha_{24}+\alpha_{42}\right)u}\frac{\sinh\left[\omega\left(\eta-u\right)\right]}{\sinh\left(\omega\eta\right)}.
\end{align}
The parameter $\eta$ is defined through the relations, 
\begin{align}
\sinh^{2}\left(\omega\eta\right) & =\frac{\omega}{\varOmega}, & \coth\left(\omega\eta\right) & =\frac{1}{\omega}\left[\alpha_{11}-\frac{1}{2}\left(\alpha_{24}+\alpha_{42}\right)\right].\label{Eta}
\end{align}

The parameters $\alpha_{ij}$ present in the expressions above denote
the derivatives of the $R$ matrix elements evaluated at zero, i.e.,
$\alpha_{ij}=r_{ij}^{\prime}(0)$. The local Hamiltonian of the models
are given by the formula $\mathcal{H}=R^{\prime}(0)P$, where $P$
denotes the permutator matrix so that $P=R(0)$ for regular $R$ matrices;
thus, we can write the Hamiltonians in the following general form:
\begin{equation}
\mathcal{H}=\left(\begin{array}{ccccccccc}
\alpha_{11} & 0 & 0 & 0 & 0 & 0 & 0 & 0 & 0\\
0 & \alpha_{24} & 0 & \alpha_{22} & 0 & 0 & 0 & 0 & 0\\
0 & 0 & \alpha_{37} & 0 & 0 & 0 & \alpha_{33} & 0 & 0\\
0 & \frac{\omega\varOmega}{\alpha_{22}} & 0 & \alpha_{42} & 0 & 0 & 0 & 0 & 0\\
0 & 0 & 0 & 0 & \alpha_{55} & 0 & 0 & 0 & 0\\
0 & 0 & 0 & 0 & 0 & \alpha_{68} & 0 & \alpha_{66} & 0\\
0 & 0 & \frac{\omega\varOmega}{\alpha_{33}} & 0 & 0 & 0 & \alpha_{24}-\alpha_{37}+\alpha_{42} & 0 & 0\\
0 & 0 & 0 & 0 & 0 & \frac{\omega\varOmega}{\alpha_{66}} & 0 & \alpha_{24}+\alpha_{42}-\alpha_{68} & 0\\
0 & 0 & 0 & 0 & 0 & 0 & 0 & 0 & \alpha_{99}
\end{array}\right),\label{Hamiltonian}
\end{equation}
provided that we have $\alpha_{99}=\alpha_{55}=\alpha_{11}$ in the
first solution; $\alpha_{55}=\alpha_{11}$ and $\alpha_{99}=-\alpha_{11}+\alpha_{24}+\alpha_{42}$
in the second solution; $\alpha_{99}=\alpha_{11}$ and $\alpha_{55}=-\alpha_{11}+\alpha_{24}+\alpha_{42}$
in the third solution and, finally, $\alpha_{99}=\alpha_{55}=-\alpha_{11}+\alpha_{24}+\alpha_{42}$
in the fourth solution. Each Hamiltonian contain eight free-parameters.

We remark that the second, third and fourth solutions are related
to each other by the similarity transformations exchanging only $r_{11}$,
$r_{55}$ and $r_{99}$; other similar solutions can also be found
through analogous similarity transformations (or by replacing $\eta$
with $-\eta$). Besides, more symmetric solutions can be found by
setting $\alpha_{42}=-\alpha_{24}$, so that the exponential factor
$\mathrm{e}^{\frac{1}{2}\left(\alpha_{24}+\alpha_{42}\right)u}$ is
removed from the expressions. This is equivalent of multiplying the
$R$ matrix by $\mathrm{e}^{-\frac{1}{2}\left(\alpha_{24}+\alpha_{42}\right)u}$
and redefining the parameters $\alpha_{ij}$. Other particular solutions
can be found by giving specific values to the parameters $\alpha_{ij}$
--- for example, the well-known fifteen-vertex $R$ matrices described
in \cite{cherednik1980method,babelon1981solutions,perk1981new} and
the solution \#4 in \cite{idzumi1994solvable} correspond to special
cases of the solutions reported here. Rational solutions can also
be obtained by taking appropriated limits. 

\section{The \boldmath $K$ matrices}

We also computed and classified the corresponding reflection $K$
matrices, solutions of the boundary Yang-Baxter equation \cite{cherednik1984factorizing,sklyanin1988boundary,mezincescu1991integrable},
\begin{equation}
R(u-v)K_{1}(u)PR(u+v)PK_{2}(v)=K_{2}(v)R(u+v)K_{1}(u)PR(u-v)P,\label{BYBE}
\end{equation}
associated with the $R$ matrices described above. We verified that
reflection $K$ matrices are the same for all those $R$ matrices.
The method employed to find these $K$ matrices is similar to the
previous one --- see Appendix \ref{AppK}.

We found that there are three families of solutions of (\ref{BYBE})
whose $K$ matrices have the following shapes:
\begin{align}
K_{1} & =\begin{pmatrix}k_{11} & 0 & 0\\
0 & k_{22} & k_{23}\\
0 & k_{32} & k_{33}
\end{pmatrix}, & K_{2} & =\begin{pmatrix}k_{11} & 0 & k_{13}\\
0 & k_{22} & 0\\
k_{31} & 0 & k_{33}
\end{pmatrix}, & K_{3} & =\begin{pmatrix}k_{11} & k_{12} & 0\\
k_{21} & k_{22} & 0\\
0 & 0 & k_{33}
\end{pmatrix}.
\end{align}
 The expressions for their elements are given in the sequence (the
parameters $\beta_{ij}$ denote the derivatives of $k_{ij}$ evaluated
at zero).

For the first solution $K_{1}$, we have: 
\begin{align}
k_{11} & =\mathrm{e}^{\beta_{11}u},\nonumber \\
k_{22} & =2\mathrm{e}^{-\frac{1}{2}\left(\alpha_{24}+2\alpha_{37}-\alpha_{42}-2\alpha_{68}-2\beta_{11}\right)u}\nonumber \\
 & \times\left[\frac{2\omega\cosh(\omega u)+\left(\alpha_{24}+\alpha_{42}-2\alpha_{68}+\beta_{22}-\beta_{33}\right)\sinh(\omega u)}{\varDelta_{1}}\right],\nonumber \\
k_{23} & =2\mathrm{e}^{-\left(\alpha_{37}-\alpha_{42}-\beta_{11}\right)u}\sinh(2\omega u)\frac{\beta_{23}}{\varDelta_{1}},\nonumber \\
k_{33} & =2\mathrm{e}^{\frac{1}{2}\left(\alpha_{24}-2\alpha_{37}+3\alpha_{42}-2\alpha_{68}+2\beta_{11}\right)u}\nonumber \\
 & \times\left[\frac{2\omega\cosh(\omega u)+\left(\alpha_{24}+\alpha_{42}-2\alpha_{68}+\beta_{22}-\beta_{33}\right)\sinh(\omega u)}{\varDelta_{1}}\right],\nonumber \\
k_{32} & =2\mathrm{e}^{-\left(\alpha_{37}-\alpha_{42}-\beta_{11}\right)u}\sinh(2\omega u)\frac{\beta_{32}}{\varDelta_{1}},
\end{align}
where, 
\begin{multline}
\varDelta_{1}=\left(2\alpha_{24}-2\alpha_{37}+\beta_{22}-\beta_{33}\right)-\left(2\alpha_{37}+2\alpha_{42}-4\alpha_{68}+\beta_{22}-\beta_{33}\right)\cosh\left(2\omega u\right)\\
-\left(2\alpha_{37}-2\alpha_{42}-2\beta_{11}+\beta_{22}+\beta_{33}\right)\sinh\left(2\omega u\right).
\end{multline}
Besides, the following constraint should be satisfied:
\begin{equation}
\beta_{23}\beta_{32}=-\left(2\alpha_{37}-2\alpha_{68}-\beta_{11}+\beta_{22}\right)\left(2\alpha_{42}-2\alpha_{68}+\beta_{11}-\beta_{33}\right).\label{beta1}
\end{equation}

For the second solution $K_{2}$, we have: 
\begin{align}
k_{11} & =2\mathrm{e}^{\frac{1}{2}\left(\alpha_{24}+2\alpha_{37}-\alpha_{42}-2\alpha_{68}+2\beta_{22}\right)u}\nonumber \\
 & \times\left[\frac{2\omega\cosh(\omega u)+\left(\alpha_{2,4}-2\alpha_{3,7}+\alpha_{4,2}+\beta_{1,1}-\beta_{3,3}\right)\sinh(\omega u)}{\varDelta_{2}}\right],\nonumber \\
k_{13} & =2\mathrm{e}^{\left(\alpha_{24}-\alpha_{68}+\beta_{22}\right)u}\sinh\left(2\omega u\right)\frac{\beta_{13}}{\varDelta_{2}},\nonumber \\
k_{22} & =\mathrm{e}^{\beta_{22}u},\nonumber \\
k_{31} & =2\mathrm{e}^{\left(\alpha_{24}-\alpha_{68}+\beta_{22}\right)u}\sinh\left(2\omega u\right)\frac{\beta_{31}}{\varDelta_{2}},\nonumber \\
k_{33} & =2\mathrm{e}^{\frac{1}{2}\left(3\alpha_{24}-2\alpha_{37}+\alpha_{42}-2\alpha_{68}+2\beta_{22}\right)u}\nonumber \\
 & \times\left[\frac{2\omega\cosh(\omega u)-\left(\alpha_{2,4}-2\alpha_{3,7}+\alpha_{4,2}+\beta_{1,1}-\beta_{3,3}\right)\sinh(\omega u)}{\varDelta_{2}}\right],
\end{align}
 where, 
\begin{multline}
\varDelta_{2}=-\left(2\alpha_{42}-2\alpha_{68}+\beta_{11}-\beta_{33}\right)+\left(2\alpha_{24}-4\alpha_{37}+2\alpha_{68}+\beta_{11}-\beta_{33}\right)\cosh\left(2\omega u\right)\\
+\left(2\alpha_{24}-2\alpha_{68}-\beta_{11}+2\beta_{22}-\beta_{33}\right)\sinh\left(2\omega u\right),
\end{multline}
and, in this case, the following constraint should be taken into account:
\begin{equation}
\beta_{13}\beta_{31}=\left(2\alpha_{37}-2\alpha_{68}-\beta_{11}+\beta_{22}\right)\left(2\alpha_{24}-2\alpha_{37}+\beta_{22}-\beta_{33}\right).\label{beta2}
\end{equation}

Finally, for the third solution $K_{3}$ we have: 
\begin{align}
k_{11} & =2\mathrm{e}^{-\frac{1}{2}\left(\alpha_{24}-2\alpha_{37}+3\alpha_{42}-2\alpha_{68}-2\beta_{33}\right)u}\nonumber \\
 & \times\left[\frac{2\omega\cosh(\omega u)-\left(\alpha_{24}-\alpha_{42}-\beta_{11}+\beta_{22}\right)\sinh(\omega u)}{\varDelta_{3}}\right],\nonumber \\
k_{12} & =2\mathrm{e}^{-\left(\alpha_{24}-\alpha_{37}+\alpha_{42}-\alpha_{68}-\beta_{33}\right)u}\sinh(2\omega u)\frac{\beta_{12}}{\varDelta_{3}},\nonumber \\
k_{21} & =2\mathrm{e}^{-\left(\alpha_{24}-\alpha_{37}+\alpha_{42}-\alpha_{68}-\beta_{33}\right)u}\sinh(2\omega u)\frac{\beta_{21}}{\varDelta_{3}},\nonumber \\
k_{22} & =2\mathrm{e}^{-\frac{1}{2}\left(3\alpha_{24}-2\alpha_{37}+\alpha_{42}-2\alpha_{68}-2\beta_{33}\right)u}\nonumber \\
 & \times\left[\frac{2\omega\cosh(\omega u)+\left(\alpha_{24}-\alpha_{42}-\beta_{11}+\beta_{22}\right)\sinh(\omega u)}{\varDelta_{3}}\right],\nonumber \\
k_{33} & =\mathrm{e}^{\beta_{33}u},
\end{align}
 where, now,
\begin{multline}
\varDelta_{3}=-\left(2\alpha_{37}-2\alpha_{68}-\beta_{11}+\beta_{22}\right)+\left(2\alpha_{24}-2\alpha_{37}-2\alpha_{42}+2\alpha_{68}-\beta_{11}+\beta_{22}\right)\cosh(2\omega u)\\
-\left(2\alpha_{24}-2\alpha_{37}+2\alpha_{42}-2\alpha_{68}+\beta_{11}+\beta_{22}-2\beta_{33}\right)\sinh(2\omega u).
\end{multline}
In this case we have the constraint:
\begin{equation}
\beta_{12}\beta_{21}=\left(2\alpha_{42}-2\alpha_{68}+\beta_{11}-\beta_{33}\right)\left(2\alpha_{24}-2\alpha_{37}+\beta_{22}-\beta_{33}\right).\label{beta3}
\end{equation}

Diagonal $K$ matrices can be obtained from the above ones by letting
the non-diagonal parameters $\beta_{ij}$ go to zero. Notice that
this forces the vanishing of the constraints (\ref{beta1}), (\ref{beta2})
and (\ref{beta3}). Thus, a priori, each case gives place to two diagonal
solutions, which would result in six possibilities in total. However,
after renormalizing the solutions (so that we always get $k_{11}=\mathrm{e}^{\beta_{11}u}$),
we verified that there are, actually, only three distinct diagonal
$K$ matrices, which are the following ones:

The first diagonal $K$ matrix is found by making either $\beta_{23}=\beta_{32}=0$
in the first solution $K_{1}$, or making $\beta_{13}=\beta_{31}=0$
in the second solution $K_{2}$, and then setting $\beta_{22}=\beta_{11}-2\left(\alpha_{37}-\alpha_{68}\right)$.
It is as follows:
\begin{align}
k_{11} & =\mathrm{e}^{\beta_{11}u},\nonumber \\
k_{22} & =\mathrm{e}^{-\left(\alpha_{24}-\alpha_{42}-\beta_{11}\right)u},\nonumber \\
k_{33} & =\mathrm{e}^{\left(\alpha_{24}-2\alpha_{37}+\alpha_{42}+\beta_{11}\right)u}\left[\frac{2\omega\cosh(\omega u)-\left(\alpha_{24}-2\alpha_{37}+\alpha_{42}+\beta_{11}-\beta_{33}\right)\sinh(\omega u)}{2\omega\cosh(\omega u)+\left(\alpha_{24}-2\alpha_{37}+\alpha_{42}+\beta_{11}-\beta_{33}\right)\sinh(\omega u)}\right].
\end{align}

The second diagonal $K$ matrix is found by making either $\beta_{23}=\beta_{32}=0$
in the first solution $K_{1}$, or making $\beta_{12}=\beta_{21}=0$
in the third solution $K_{3}$, and then setting $\beta_{33}=\beta_{11}+2\left(\alpha_{42}-\alpha_{68}\right)$.
In this case we get: 
\begin{align}
k_{11} & =\mathrm{e}^{\beta_{11}u},\nonumber \\
k_{22} & =\mathrm{e}^{-\left(\alpha_{24}-\alpha_{42}-\beta_{11}\right)u}\left[\frac{2\omega\cosh(\omega u)+\left(\alpha_{24}-\alpha_{42}-\beta_{11}+\beta_{22}\right)\sinh(\omega u)}{2\omega\cosh(\omega u)-\left(\alpha_{24}-\alpha_{42}-\beta_{11}+\beta_{22}\right)\sinh(\omega u)}\right].\nonumber \\
k_{33} & =\mathrm{e}^{\left(\alpha_{24}-2\alpha_{37}+\alpha_{42}+\beta_{11}\right)u};
\end{align}

Finally, the third diagonal $K$ matrix is found by making either
$\beta_{13}=\beta_{31}=0$ in the second solution $K_{2}$, or making
$\beta_{12}=\beta_{21}=0$ in the third solution $K_{3}$, and then
setting $\beta_{33}=\beta_{22}+2\left(\alpha_{24}-\alpha_{37}\right)$.
This results in the following:
\begin{align}
k_{11} & =\mathrm{e}^{\beta_{11}u},\nonumber \\
k_{22} & =\mathrm{e}^{-\left(\alpha_{24}-\alpha_{42}-\beta_{11}\right)u}\left[\frac{2\omega\cosh(\omega u)+\left(\alpha_{24}-\alpha_{42}-\beta_{11}+\beta_{22}\right)\sinh(\omega u)}{2\omega\cosh(\omega u)-\left(\alpha_{24}-\alpha_{42}-\beta_{11}+\beta_{22}\right)\sinh(\omega u)}\right],\nonumber \\
k_{33} & =\mathrm{e}^{\left(\alpha_{24}-2\alpha_{37}+\alpha_{42}+\beta_{11}\right)u}\left[\frac{2\omega\cosh(\omega u)+\left(\alpha_{24}-\alpha_{42}-\beta_{11}+\beta_{22}\right)\sinh(\omega u)}{2\omega\cosh(\omega u)-\left(\alpha_{24}-\alpha_{42}-\beta_{11}+\beta_{22}\right)\sinh(\omega u)}\right].
\end{align}

The reflection $K$ matrices above generalize the respective diagonal
$K$ matrices derived in \cite{Vega1993boundary} and the non-diagonal
ones found in \cite{Lima2002}.

\section{Conclusion}

In this work, we classified all the regular $R$ and $K$ matrices
of fifteen-vertex models whose $R$ matrix has the usual shape (\ref{R}).
This represents the first step towards the classification of the regular
$R$ and $K$ matrices associated with spin-1 (three-states) vertex
models. The $R$ matrices were obtained through an algebraic-differential
method developed in \cite{vieira2018solving}. We believe that this
method is powerful enough to solve the Yang-Baxter equation for spin-1
vertex models with different initial shapes for the $R$ matrices.
This could lead to new solutions to the Yang-Baxter equation. As a
simple example, it is straightforward to verify from the algebraic-differential
method that the most general regular solution of the Yang-Baxter equation
for \emph{nine-vertex models} has the form of a ``dressed'' permutation
matrix, i.e., the elements of this $R$ matrix are as follows:
\begin{equation}
r_{ij}=\mathrm{e}^{\alpha_{ij}}P_{ij}.
\end{equation}
Moreover, we comment in advance that we already succeeded in finding
non-symmetric solutions of the Yang-Baxter equation for nineteen-vertex
models, which generalize the famous Zamolodchikov-Fateev and Izergin-Korepin
vertex models. These solutions will be reported soon.

An interesting question to be worked out is the implementation of
the algebraic Bethe Ansatz for these fifteen-vertex models. This would
require a generalization of the nested Bethe Ansatz to take account
the asymmetry of the $R$ matrices weights.

\appendix

\section{Solving the periodic Yang-Baxter equation\label{AppR}}

The $R$ matrices presented in this work were obtained through an
algebraic-differential approach developed in \cite{vieira2018solving},
where all the regular, $Z_{2}$ symmetric, $R$ matrices for spin-1/2
(two-states) vertex models were classified --- see also \cite{sogo1982classification,khachatryan2013solutions}
for previous classifications and \cite{deLeew2019classifying} for
solutions without $Z_{2}$ symmetry). In short, in this method, instead
of solving directly the system of functional equations that the Yang-Baxter
equation (\ref{YBE}) represents --- which would be very hard ---,
we turn our attention to the following equivalent systems of differential
equations:
\begin{align}
U & \coloneqq R_{12}(u)D_{13}(u)P_{23}+R_{12}(u)R_{13}(u)H_{23}=H_{23}R_{13}(u)R_{12}(u)+P_{23}D_{13}(u)R_{12}(u),\label{U}\\
V & \coloneqq R_{23}(v)D_{13}(v)P_{12}+R_{23}(v)R_{13}(v)H_{12}=H_{12}R_{13}(v)R_{23}(v)+P_{12}D_{13}(v)R_{23}(v),\label{V}
\end{align}
where, 
\begin{align}
D(u) & =\left.\frac{\partial R(u+v)}{\partial v}\right|_{v=0}, & D(v) & \left.\frac{\partial R(u+v)}{\partial u}\right|_{u=0}, & P & =R(0), &  & \text{and} & H & =D(0).\label{DD}
\end{align}

Equations (\ref{U}) and (\ref{V}) are obtained by differentiating
(\ref{YBE}) with respect to the variables $u$ and $v$, respectively,
and then evaluating these derivatives at zero. The derivatives ($d_{ij}$)
of the $R$ matrix elements ($r_{ij}$), however, are regarded as
independent variables, so that we can say that the original system
of functional equations (\ref{YBE}) is actually replaced by two systems
of \emph{algebraic equations} for the unknowns $r_{ij}$ and $d_{ij}$.
In this way, we can verify that systems (\ref{U}) and (\ref{V})
can be completely solved by algebraic means, although in general not
all elements of the $R$ matrix can be determined from these equations
(the reason is that these systems of equations are usually overdetermined).
The remaining elements of the $R$ matrix can, notwithstanding, be
determined from the consistency conditions $d_{ij}=r_{ij}^{\prime}$,
which give place to a few number of simple differential equations.
See \cite{vieira2018solving} for more details.

The idea of transforming a functional equation into a differential
one seems to be first considered by Abel \cite{Abel1823,Aczel1966},
the great Norwegian mathematician. Regarding the Yang-Baxter equation,
we remark that some of the first solutions of the Yang-Baxter equations
were found by a similar differential approach \cite{KulishSklyanin1982},
although only symmetric $R$ matrices were considered (certainly due
to the lack of computational power available at the time) and no systematic
analysis of the equations were considered. We highlight that the algebraic-differential
method developed in \cite{vieira2018solving} and used here has several
advantages when compared with other approaches. Firstly, it is usually
simpler. Secondly, it is a powerful method, which allows one to solve
the Yang-Baxter equation without imposing any constraints on the $R$
matrix elements. Besides, the existence and unicity of the solutions
can be ensured directly from the underlining theory of differential
equations, as well as, the generality of the solutions can be guaranteed
from the analysis of the branches of the algebraic equations ---
for instance, using the tools of algebraic geometry. Finally, the
method provides straightforwardly the Hamiltonian $\mathcal{H}$ of
the integrable models through the simple expression: $\mathcal{H}=HP$,
where $H$ is the derivative of the $R$ matrix evaluated at zero.

In the following, we shall discuss in detail how the systems of equations
(\ref{U}) and (\ref{V}) were solved. We assumed that the non-null
elements of the $R$ matrix (\ref{R}) are always different from zero.
In the expressions below, $\alpha_{ij}$ denote the derivatives of
$r_{ij}$ evaluated at zero.

A first analysis of equations (\ref{U}) and (\ref{V}) shows us the
presence of several simple relations involving only the diagonal elements
of the $R$ matrix and their derivatives. Thus, we can eliminate several
unknowns in a straightforward way. For example, from the pairs of
equations $\left\{ U_{6,8},V_{6,8}\right\} $, $\left\{ U_{12,16},V_{12,16}\right\} $,
$\left\{ U_{6,12},V_{6,12}\right\} $, $\left\{ U_{22,20},V_{22,20}\right\} $
and $\left\{ U_{8,20},V_{8,20}\right\} $ we can find, respectively,
the expressions for $r_{33}$, $r_{44}$, $r_{66}$, $r_{77}$, $r_{88}$
and their derivatives. They are given by simple expressions depending
only on $r_{22}$ or $d_{22}$: 
\begin{align}
r_{33} & =\left(\frac{\alpha_{33}}{\alpha_{22}}\right)r_{22}, & r_{44} & =\left(\frac{\alpha_{44}}{\alpha_{22}}\right)r_{22}, & r_{66} & =\left(\frac{\alpha_{66}}{\alpha_{22}}\right)r_{22},\nonumber \\
r_{77} & =\left(\frac{\alpha_{77}}{\alpha_{22}}\right)r_{22}, & r_{88} & =\left(\frac{\alpha_{88}}{\alpha_{22}}\right)r_{22},
\end{align}
\begin{align}
d_{33} & =\left(\frac{\alpha_{33}}{\alpha_{22}}\right)d_{22}, & d_{44} & =\left(\frac{\alpha_{44}}{\alpha_{22}}\right)d_{22}, & d_{66} & =\left(\frac{\alpha_{66}}{\alpha_{22}}\right)d_{22},\nonumber \\
d_{77} & =\left(\frac{\alpha_{77}}{\alpha_{22}}\right)d_{22}, & d_{88} & =\left(\frac{\alpha_{88}}{\alpha_{22}}\right)d_{22}.
\end{align}

Next, we look for the equations containing only non-diagonal elements
of the $R$ matrix and their derivatives. It can be verified that
equations $U_{4,4}$, $U_{7,7}$ and $U_{17,17}$, for example, become
equivalent to the following: 
\begin{align}
\frac{d_{24}}{r_{24}}-\frac{d_{42}}{r_{42}} & =\alpha_{24}-\alpha_{42}, & \frac{d_{37}}{r_{37}}-\frac{d_{73}}{r_{73}} & =\alpha_{37}-\alpha_{73}, & \frac{d_{68}}{r_{68}}-\frac{d_{86}}{r_{86}} & =\alpha_{68}-\alpha_{86}.
\end{align}
Thus, if we integrate these equations using of the initial conditions
$r_{42}(0)/r_{24}(0)=r_{73}(0)/r_{37}(0)=r_{86}(0)/r_{68}(0)=1$,
then the following relations would be obtained: 
\begin{align}
\frac{r_{24}}{r_{42}} & =\mathrm{e}^{\left(\alpha_{24}-\alpha_{42}\right)u}, & \frac{r_{37}}{r_{73}} & =\mathrm{e}^{\left(\alpha_{37}-\alpha_{73}\right)u}, & \frac{r_{68}}{r_{86}} & =\mathrm{e}^{\left(\alpha_{68}-\alpha_{86}\right)u}.
\end{align}
From this we could eliminate, if we wish, the unknowns $r_{42}$,
$r_{73}$ and $r_{86}$ in terms of $r_{24}$, $r_{37}$ and $r_{68}$,
respectively. However, to avoid introducing exponentials into the
systems of equations --- which would make them harder ---, we shall
not proceed in this way. Instead, we shall use equations $U_{4,4}$,
$U_{7,7}$ and $U_{17,17}$ just to eliminate the derivatives $d_{42}$,
$d_{73}$ and $d_{86}$, which become: 
\begin{align}
d_{42} & =-\left(\alpha_{24}-\alpha_{42}\right)r_{42}+\left(\frac{r_{42}}{r_{24}}\right)d_{24},\nonumber \\
d_{73} & =-\left(\alpha_{37}-\alpha_{73}\right)r_{73}+\left(\frac{r_{73}}{r_{37}}\right)d_{37},\nonumber \\
d_{86} & =-\left(\alpha_{68}-\alpha_{86}\right)r_{86}+\left(\frac{r_{86}}{r_{68}}\right)d_{68}.
\end{align}
 Now, equations $V_{8,22}$, $U_{6,20}$ and $V_{6,30}$ provide the
expressions for $d_{2,4}$, $d_{3,7}$ and $d_{6,8}$: 
\begin{align}
d_{24} & =\left(\alpha_{37}-\alpha_{68}\right)r_{24}+\left(\frac{r_{24}}{r_{22}}\right)d_{22}-\left(\frac{r_{37}r_{86}}{r_{22}}\right)\alpha_{22},\nonumber \\
d_{37} & =-\left(\alpha_{42}-\alpha_{68}\right)r_{37}+\left(\frac{r_{37}}{r_{22}}\right)d_{22}-\left(\frac{r_{24}r_{68}}{r_{22}}\right)\alpha_{22},\nonumber \\
d_{68} & =-\left(\alpha_{24}-\alpha_{37}\right)r_{68}+\left(\frac{r_{68}}{r_{22}}\right)d_{22}-\left(\frac{r_{37}r_{42}}{r_{22}}\right)\alpha_{22},
\end{align}
and, then, from equations $U_{2,4}$, $V_{5,11}$ and $V_{9,21}$,
we can eliminate $d_{11}$, $d_{55}$ and $d_{99}$:
\begin{align}
d_{11} & =\left(\frac{r_{11}}{r_{22}}\right)d_{22}-\left(\frac{r_{24}r_{42}}{r_{22}}\right)\alpha_{22},\nonumber \\
d_{55} & =\left(\frac{r_{55}}{r_{22}}\right)d_{22}-\left(\frac{r_{24}r_{42}}{r_{22}}\right)\alpha_{22},\nonumber \\
d_{99} & =\left(\frac{r_{99}}{r_{22}}\right)d_{22}-\left(\frac{r_{37}r_{73}}{r_{22}}\right)\alpha_{22}.
\end{align}
After that, equations $U_{3,7}$ and $U_{15,17}$ give simple relations
for $r_{73}$ and $r_{86}$: 
\begin{align}
r_{73} & =\frac{r_{24}r_{42}}{r_{37}}, & r_{86} & =\frac{r_{24}r_{42}}{r_{68}}.
\end{align}
 Next, we use equation $F_{6,16}$ to eliminate $d_{22}$: 
\begin{equation}
d_{22}=-\left(\alpha_{24}-\alpha_{86}\right)r_{22}+\left(\frac{d_{37}}{r_{37}}\right)r_{22}+\left(\frac{r_{24}r_{68}}{r_{37}}\right)\alpha_{22},
\end{equation}
 and from equations $U_{4,2}$, $U_{5,11}$ and $U_{9,21}$, we can
eliminate $r_{11}$, $r_{55}$ and $r_{99}$:
\begin{align}
r_{11} & =\frac{r_{37}r_{42}}{r_{68}}+\left(\frac{\alpha_{11}-\alpha_{37}-\alpha_{42}+\alpha_{68}}{\alpha_{22}}\right)r_{22},\nonumber \\
r_{55} & =\frac{r_{37}r_{42}}{r_{68}}+\left(\frac{-\alpha_{37}-\alpha_{42}+\alpha_{55}+\alpha_{68}}{\alpha_{22}}\right)r_{22},\nonumber \\
r_{99} & =\frac{r_{24}r_{68}}{r_{37}}+\left(\frac{-\alpha_{24}+\alpha_{37}-\alpha_{68}+\alpha_{99}}{\alpha_{22}}\right)r_{22}.
\end{align}
Finally, from $V_{3,7}$ we eliminate $r_{42}$: 
\begin{equation}
r_{42}=\left(\frac{r_{68}}{r_{37}}\right)^{2}r_{24}+\left(\frac{-\alpha_{24}+\alpha_{37}+\alpha_{42}-\alpha_{68}-\alpha_{73}+\alpha_{86}}{\alpha_{22}}\right)\frac{r_{22}r_{68}}{r_{37}}.
\end{equation}

At this point, several constraints between the parameters $\alpha_{ij}$
emerge, which means that some of them must be fixed in terms of the
others in order to equations (\ref{U}) and (\ref{V}) be satisfied.
For example, from equations $U_{8,12}$ and $V_{12,8}$ we can fix
$\alpha_{73}$ and $\alpha_{86}$: 
\begin{align}
\alpha_{73} & =\alpha_{24}-\alpha_{37}+\alpha_{42}, & \alpha_{86} & =\alpha_{24}+\alpha_{42}-\alpha_{68}.
\end{align}
Then, from $U_{2,10}$, $U_{6,22}$ and $U_{12,20}$, we can also
fix the parameters $\alpha_{44}$, $\alpha_{77}$ and $\alpha_{88}$:
\begin{align}
\alpha_{44} & =\left.\left(\alpha_{11}-\alpha_{24}+\alpha_{37}-\alpha_{68}\right)\left(\alpha_{11}-\alpha_{37}-\alpha_{42}+\alpha_{68}\right)\right/\alpha_{22},\nonumber \\
\alpha_{77} & =\left.\left(\alpha_{11}-\alpha_{24}+\alpha_{37}-\alpha_{68}\right)\left(\alpha_{11}-\alpha_{37}-\alpha_{42}+\alpha_{68}\right)\right/\alpha_{33},\nonumber \\
\alpha_{88} & =\left.\left(\alpha_{11}-\alpha_{24}+\alpha_{37}-\alpha_{68}\right)\left(\alpha_{11}-\alpha_{37}-\alpha_{42}+\alpha_{68}\right)\right/\alpha_{66}.
\end{align}
Now we can verify that all the remaining equations will be satisfied
whenever one of the following two additional constraints are imposed:
\begin{align}
\left(\alpha_{11}-\alpha_{55}\right)\left(\alpha_{11}-\alpha_{24}-\alpha_{42}+\alpha_{55}\right) & =0, &  & \text{or} & \left(\alpha_{11}-\alpha_{99}\right)\left(\alpha_{11}-\alpha_{24}-\alpha_{42}+\alpha_{99}\right) & =0.\label{Branches}
\end{align}
This leads to four families of solutions according to what factor
we choose to vanish in each of the relations above. The possibilities
are the following:
\begin{itemize}
\item If $\alpha_{55}=\alpha_{11}$ and $\alpha_{99}=\alpha_{11}$, then
we get a solution in which $r_{99}=r_{55}=r_{11}$ (the first solution);
\item If $\alpha_{55}=\alpha_{11}$ and $\alpha_{99}=-\alpha_{11}+\alpha_{24}+\alpha_{42}$,
then we get a solution in which $r_{55}=r_{11}$ only (the second
solution);
\item If $\alpha_{55}=-\alpha_{11}+\alpha_{24}+\alpha_{42}$ and $\alpha_{99}=\alpha_{11}$,
then we get a solution in which $r_{99}=r_{11}$ only (the third solution);
\item If $\alpha_{55}=-\alpha_{11}+\alpha_{24}+\alpha_{42}$ and $\alpha_{99}=-\alpha_{11}+\alpha_{24}+\alpha_{42}$,
then we get a solution in which $r_{99}=r_{55}$ only (the fourth
solution).
\end{itemize}

Here we should notice that, although all the equations of systems
(\ref{U}) and (\ref{V}) are already satisfied when the constraints
given above are taken into account, it still remains to find the expressions
for $r_{22}$, $r_{24}$, $r_{68}$, $r_{37}$ and $d_{37}$. This
happens because systems (\ref{U}) and (\ref{V}) are overdetermined
--- i.e., they have a positive Hilbert dimension, in the jargon of
algebraic geometry. Nonetheless, the expressions for the remaining
unknowns can be found by imposing, for consistency, that $d_{ij}$
are indeed the derivatives of $r_{ij}$. Therefore, from the previously
obtained expressions for $d_{22}$, $d_{24}$ and $d_{68}$, we get
the following simple system of differential equations:
\begin{align}
r_{22}^{\prime} & =\left(\alpha_{42}-\alpha_{68}\right)r_{22}+\frac{r_{24}r_{68}}{r_{37}}\left(\frac{d_{37}}{r_{37}}\right)\alpha_{22}r_{22},\nonumber \\
r_{24}^{\prime} & =\left(\frac{d_{37}}{r_{37}}+\alpha_{24}-\alpha_{37}\right)r_{24},\nonumber \\
r_{68}^{\prime} & =\left(\frac{d_{37}}{r_{37}}-\alpha_{37}+\alpha_{68}\right)r_{68}.\label{EDO}
\end{align}
To simplify further these equations, we can use the fact that the
expression for $r_{37}$ can be chosen as any function $f(u)$ that
satisfies the relations $f(0)=1$ and $f^{\prime}(0)=\alpha_{37}$.
Thus, we can choose:
\begin{align}
r_{37} & =\mathrm{e}^{\alpha_{37}u}, & d_{37} & =\alpha_{37}\mathrm{e}^{\alpha_{37}u},
\end{align}
so that (\ref{EDO}) becomes:
\begin{align}
r_{22}^{\prime} & =\mathrm{e}^{-\alpha_{37}u}\alpha_{22}r_{24}r_{68}+\left(\alpha_{37}+\alpha_{42}-\alpha_{68}\right)r_{22}, & r_{24}^{\prime} & =\alpha_{24}r_{24}, & r_{68}^{\prime} & =\alpha_{68}r_{68}.
\end{align}
This system can be easily solved for the initial conditions $r_{22}(0)=0$,
$r_{24}(0)=1$ and $r_{68}(0)=1$. The solution is: 
\begin{align}
r_{22} & =\left(\frac{\alpha_{22}}{\omega}\right)\mathrm{e}^{\frac{1}{2}\left(\alpha_{24}+\alpha_{42}\right)u}\sinh\left[\frac{1}{2}\left(\alpha_{24}-2\alpha_{37}-\alpha_{42}+2\alpha_{68}\right)u\right], & r_{24} & =\mathrm{e}^{\alpha_{24}u}, & r_{68} & =\mathrm{e}^{\alpha_{68}u}.
\end{align}

The $R$ matrices presented in the main text follow after we define
the quantities $\omega$, $\varOmega$ and $\eta$ as given by (\ref{Omegas})
and (\ref{Eta}) and simplify all the expressions.

\section{Solving the boundary Yang-Baxter equation\label{AppK}}

A similar method can be employed to solve the boundary Yang-Baxter
equation (\ref{BYBE}). In fact, the derivatives of (\ref{BYBE})
with respect to $u$ and $v$ evaluated at zero provide, respectively,
the following systems of algebraic equations:
\begin{multline}
\mathcal{U}\coloneqq-B_{2}R(u)K_{1}(u)PR(u)P+R(u)K_{1}(u)PR(u)PB_{2}\\
-2D(u)K_{1}(u)PR(u)P+2R(u)K_{1}(u)PD(u)P=0,\label{UU}
\end{multline}
 and 
\begin{multline}
\mathcal{V}\coloneqq-K_{2}(v)R(v)B_{1}PR(-v)P+R(-v)B_{1}PR(v)PK_{2}(v)+D(-v)PR(v)PK_{2}(v)\\
-K_{2}(v)D(v)PR(-v)P-K_{2}(v)R(v)PD(-v)P+R(-v)PD(v)PK_{2}(v)=0.\label{VV}
\end{multline}
where the matrices $R$, $D$ and $P$ are the same as defined in
(\ref{DD}), while 
\begin{equation}
K=\begin{pmatrix}k_{11} & k_{12} & k_{13}\\
k_{21} & k_{22} & k_{23}\\
k_{31} & k_{32} & k_{33}
\end{pmatrix},\label{K}
\end{equation}
 is the reflection matrix and $B$ corresponds to the derivative of
the $K$ matrix evaluated at zero: 
\begin{align}
B & =\left.\frac{\mathrm{d}K(u)}{\mathrm{d}u}\right|_{u=0}=\left.\frac{\mathrm{d}K(v)}{\mathrm{d}v}\right|_{v=0}.
\end{align}
We also made use of the regularity condition $K(0)=I$, where $I$
is the identity matrix. This method for solving the boundary Yang-Baxter
equation is very well-known. Thus, it is not necessary to give a detailed
exposition about how equations (\ref{UU}) and (\ref{VV}) are solved,
so that in the what follows we shall present only the main steps.

Let $k_{ij}$ denote the elements of the $K$ matrix and $\beta_{ij}$
their derivatives evaluated at zero. Inserting the $K$ matrix (\ref{K})
into (\ref{UU}) and (\ref{VV}) and simplifying, we can verify in
a first analysis that these systems of equations are only consistent
if either $k_{12}=0$ or $k_{13}=0$ or both are zero. Besides, if
any of $k_{12}$ or $k_{13}$ is zero, then we get as well that the
respective transpose, $k_{21}$ or $k_{31}$, should also be zero.
Thus, we are led to the following: if both $k_{12}$ and $k_{13}$
are zero, then we shall get solution named $K_{1}$ in the main text;
if $k_{12}=0$ but $k_{13}\neq0$, then we shall get solution $K_{2}$;
finally, if $k_{12}\neq0$ but $k_{13}=0$ then we shall get solution
$K_{3}$. In any case, the recipe to find the remaining elements of
the $K$ matrices is the same: we first eliminate the non-diagonal
elements of the $K$ matrix in terms of a given pivotal element (we
have chosen this pivot as $k_{11}$ for the solution $K_{1}$, as
$k_{22}$ for the solution $K_{2}$ and as $k_{33}$ for the solution
$K_{3}$). Then, the other diagonal elements can also be eliminated
in terms of the chosen pivot. Finally, we have chosen, for the solutions
$K_{1}$, $K_{2}$ and $K_{3}$ respectively, the following usual
expressions for the the pivots: $k_{11}=\mathrm{e}^{\beta_{11}u}$,
$k_{22}=\mathrm{e}^{\beta_{22}u}$ and $k_{33}=\mathrm{e}^{\beta_{33}u}$.
This fixes all the elements of the $K$ matrices. Nonetheless, the
system of equations (\ref{UU}) and (\ref{VV}) are not yet satisfied;
to this end it is necessary to impose further one simple constraint
between the non-diagonal parameters $\beta_{ij}$ of each solution
--- these are the respective constraints presented in the main text
by equations (\ref{beta1}), (\ref{beta2}) and (\ref{beta3}).

\bibliographystyle{JHEP}
\bibliography{DYBE-15V}

\end{document}